\documentclass[12pt]{article}
\usepackage{amsmath, amsthm, amssymb}

\usepackage{epsfig}
\usepackage{graphicx}
\usepackage{epstopdf}
\usepackage{subfigure}
\usepackage{fancyhdr} 

\advance\textheight by \topskip
\begin{document}

\clearpage
\hoffset -0.15cm 
\tolerance=100000
\thispagestyle{empty}
\setcounter{page}{1}

\begin{flushright}
{SHEP-08-23}\\
\today
\end{flushright}
\begin{center}
{\Large {\bf Higgs Production in Association with \\[0.15cm]
Top Squark Pairs in the MSSM at the LHC: \\[0.35cm]
the Decay Patterns\footnote{Work supported in 
part by the U.K. Science and Technology Facilities Council 
(STFC) and
by the European Union (EU) under contract MRTN-CT-2006-035505 (HEPTOOLS FP6 RTN).}}}
\\[1.5cm]
{\large H.F. Heath$^{1}$, C. Lynch$^{1,2}$, S. Moretti$^{2,3}$ \\[0.15cm]
and C.H. Shepherd-Themistocleous$^{2}$}
\\[0.5cm]
{\it $^1${H. H. Wills Physics Laboratory, Tyndall Avenue, Bristol BS8 1TL, UK}}
\\[0.5cm]
{\em $^2$Particle Physics Department, Rutherford Appleton Laboratory, Chilton, Didcot, Oxon OX11 0QX, UK}
\\[0.5cm]
{\it $^3$School of Physics and Astronomy, University of Southampton}\\
{\it Highfield, Southampton SO17 1BJ, UK}
\\[0.25cm]
\end{center}

\begin{abstract}
\noindent
The production of the lightest Higgs boson, 
$h^0$, in association 
with the lightest stop, $\tilde{t}_{1}$, 
in  the Minimal Supersymmetric Standard Model,
is an interesting channel, as its cross section can be higher than Higgs production in association with top quarks.  
Furthermore, the $\tilde{t}_{1}\tilde{t}^{*}_{1}h^{0}$ production rate is highly dependent on various Supersymmetric parameters. 
The mass, mixings, couplings and production cross sections relevant
to the $\tilde{t}_{1}\tilde{t}^{*}_{1}h^{0}$ channel have been studied in the past.
Here, we complement these analyses by performing a thorough decay study. We conclude that there is some scope for extracting this channel
at the Large Hadron Collider, for suitable combinations of the Supersymmetric parameters, in several different decay channels of both the stop quarks and Higgs boson,
the most numerically promising being the signature involving charm 
decays of the stops and bottom decay of the Higgs. This, in fact, remains
true in the Supergravity inspired Minimal Supersymmetric
Standard Model.
\end{abstract}

\pagebreak

\section*{Introduction and motivations}

The Standard Model (SM) of particle physics has been extremely successful at explaining experimental data to date.
It has, however, failed to provide answers to fundamental problems, such as the unification of the gauge couplings at high energies, as well as
the difference between such a unification scale and that of Electro-Weak (EW)
interactions (known as the hierarchy problem). 
Introducing a new fundamental symmetry of Nature between bosons and fermions, known as Supersymmetry (SUSY), solves both these problems.
If SUSY exists at the EW scale, it should be discovered at the Large Hadron Collider (LHC).
In the Minimal Supersymmetric Standard Model (MSSM), all known particles will have Supersymmetric partners, known as sparticles. 
Additionally, in the MSSM, two
complex Higgs doublets are required to avoid anomalies, which in turn have
Supersymmetric partner states. EW Symmetry Breaking (EWSB) results in five Higgs bosons in the MSSM.
The lightest neutral Higgs state of the MSSM, $h^{0}$, is likely to have a mass of less than 135 GeV, so it is well within
the kinematic reach of the LHC. Current experimental limits on the mass of the sparticles that could enter $h^0$ decay channels are such
that (with the possible exception of invisible $h^0$ decays, e.g., into two neutralinos, the SUSY partners of the SM neutral gauge/Higgs bosons,
which are however small \cite{gg,VBF,HS}) the $h^0$ decay modes that are interesting phenomenologically are those involving
SM particles (henceforth we will refer to these as SM-like channels, likewise for the production modes involving
couplings to SM particles only). However, it does not follow that $h^0$
decay modes and possible discovery channels should be the same as for a low mass SM Higgs state (hereafter denoted by $H$).
In fact, Supersymmetric effects can affect both the Higgs production and decay rates.

The $\tilde{t}_{1}\tilde{t}^{*}_{1}h^{0}$ production channel is interesting for two main reasons: 1.
it provides a production mechanism for the lightest Supersymmetric Higgs boson, in addition or as an alternative to the SM-like channels;
2. its production rates are strongly dependent on the MSSM parameters.
For $M_{h^0}\lesssim$ 135 GeV, the dominant decay mode is $h^{0} \rightarrow b\bar{b}$, however, at the LHC the inclusive $h^{0} \rightarrow b\bar{b}$ decay is not visible due to the large QCD background.
The  $h^{0} \rightarrow \gamma \gamma$ decay mode is the most probable LHC discovery channel 
for a Higgs boson at these masses \cite{hgamma}. The $h^{0} \rightarrow \gamma \gamma$ decay,
however, can be affected by Supersymmetric effects \cite{susyhiggs}, in such
a way that cancellations between the SM particle loops and those involving their SUSY partners can occur over large regions of the MSSM
parameter space. This may deplete the $h^0\to \gamma\gamma$ decay rates below experimental observability\footnote{This
applies not only to the exclusive $\gamma\gamma\ell^\pm \nu_\ell$ channel but even more so to the inclusive
mode, as a similar mechanism onsets in the suppression of the $gg\to h^0$ production channel.}.
This typically occurs for light stop quarks, with a mass
of order of that of the top quark or below. In turn, this means that one could re-establish the lost SM-like di-photon signal by exploiting
Higgs production in association with (light) stop quarks.
Even when  associated Higgs production and decay has a lower cross section than the inclusive di-photon channel, 
the extra final state particles can render this SUSY channel visible over the background by providing useful triggers, thereby possibly also 
enabling $h^0\to b\bar b$ detection.

Under the above circumstances, one can elaborate on point 1 as follows. 
Take for example the SM counterpart of  $\tilde{t}_{1}\tilde{t}^{*}_{1}h^{0}$, i.e.,
Higgs boson production in association with top quarks,  ${t}\bar t h$. This channel has been studied by both ATLAS \cite{ATLtthnote} and 
 CMS  \cite{CMStthnote}. Here, the associated top quarks decay to $b$-quarks and one of the $t$-quarks decays leptonically and the other hadronically. 
Since the dominant decay of the $H$ state is $h \rightarrow b\bar{b}$, a four $b$-quark final state is produced.  
This is much more visible over the QCD background than the inclusive two $b$-quark final state. 
The scope of this channel in the SM is now questionable though, as its feasibility is not certain \cite{CMStthnote}. 
However, in the MSSM, when there is a large stop mixing, the coupling of the lightest stop to the lightest Higgs boson can be larger than the top Yukawa coupling.
This means that in the MSSM the $\tilde{t}_{1}\tilde{t}^{*}_{1}h^{0}$ cross section can be larger than the $t\bar{t}h$ one.
In addition, for stop masses suitably larger than $m_t$, the $\tilde{t}_{1}\to t\tilde\chi_{1}^0$ decay could occur with a Branching Ratio (BR) nearing one.
The final outcome would then be a signature similar to that emerging in the SM from $t\bar t h$, possibly with larger rates and an additional handle for background
rejection, the larger missing transverse energy (due to the two neutralinos escaping detection) with respect to the SM case (when it is only due to one neutrino
produced in the leptonic top decay). 
The $\tilde{t}_{1}\tilde{t}^{*}_{1}h^{0}$ channel therefore has the potential to be an interesting channel for Higgs discovery in the MSSM and this is the reason
why it has received considerable attention in the literature \cite{moretti,djou}.  

The $\tilde{t}_{1}\tilde{t}^{*}_{1}h^{0}$ channel can also provide information on the underlying Supersymmetric scenario, point 2 above.
If SUSY exists at the EW scale, particularly in the form of light squarks, it should be discovered at the LHC in direct $\tilde{t}_{1}\tilde{t}^{*}_1$
production (i.e., without subsequent $h^0$ emission).  
However, information on the particular Supersymmetric model and its parameters will be hard to determine from this process (primarily because
it proceeds via QCD interactions, which are not modified with respect to the SM by SUSY, so that the only SUSY parameter measurable is here $m_{\tilde t_1}$). 
The $\tilde{t}_{1}\tilde{t}^{*}_{1}h^{0}$ cross section, however, is dependent on several low energy MSSM parameters.
These enter the $\tilde{t}_{1}\tilde{t}^{*}_{1}h^{0}$ cross section through both the lightest stop mass and its couplings to the lightest Higgs boson.
Measuring the $\tilde{t}_{1}\tilde{t}^{*}_{1}h^{0}$ cross section could therefore provide us with decisive information on the SUSY-breaking scenario and its parameters.  
Importantly, such low energy SUSY parameters can directly be related to those defining  the Supergravity inspired MSSM (MSUGRA),
which is  one of the simplest possible scenarios in the choice of the soft SUSY-breaking parameters at the Grand Unification (GUT) scale.
In MSUGRA, the scalar masses, scalar couplings and gaugino masses are all equal (separately) at the GUT scale.
The remaining two parameters needed to describe soft SUSY breaking in MSUGRA are the soft Higgs mixing parameter, $b$, and the Higgs mixing parameter of the 
two doublets, $\mu$.
However, using constraints from the known value of the $Z$ mass, these two parameters can be calculated from two EW scale parameters: $\tan\beta$ 
(the ratio of the Vacuum Expectation Values (VEVs) of the two Higgs doublet fields)
and the sign of $\mu$.
This means that the over one hundred MSSM parameters can be reduced to the following five MSUGRA ones. 

1. $M_{0}$, the universal scalar mass, defined at the GUT scale.

2. $M_{\frac{1}{2}}$, the universal gaugino mass, defined at the GUT scale.

3. $A_{0}$, the universal trilinear coupling, defined at the GUT scale.

4. $\tan\beta$, defined at the EW scale.

5. ${\rm{sign}}(\mu)$, the sign of the Higgs mass parameter, defined at the EW scale.

\noindent 
The $\tilde{t}_{1}\tilde{t}^{*}_{1}h^{0}$ cross section depends at lowest order on $\tan\beta$ and ${\rm{sign}}(\mu)$ defined at the EW scale
and -- through the Renormalisation Group Evolution (RGE) equations of the model --
also on the $M_0$, $M_{\frac{1}{2}}$ and  $A_0$ GUT parameters. 

In this paper we examine possible decay channels 
and test their suitability for detection at the LHC. 
In the next section, we review existing literature and update old results by using 
more recent simulation software. We also perform new scans over the MSSM and MSUGRA parameter spaces, where both
production and
decay rates are determined. Finally, we conclude in the last section. 

\section*{The $\tilde{t}_{1}\tilde{t}^{*}_{1}h^{0}$ rates in the MSUGRA parameter space}

Since the publication of \cite{moretti,djou}, the experimental constraints have been revised, and the theoretical models have been improved.
In this, and the following, section we describe the updates to these results, using the latest simulation packages, primarily those
that determine the MSSM and MSUGRA spectra of masses, couplings and BRs. Here we use
ISASUSY and ISASUGRA \cite{isa} and both were
interfaced to the Supersymmetric
version \cite{SHERWIG} of the HERWIG Monte Carlo (MC) program \cite{HERWIG}, where the production process (tested against FeynHiggs \cite{feynhiggs})
is implemented alongside the (matrix element modelled) decays of the final state objects.

There are previous papers that describe studies of Higgs production in association with light squark pairs.
The first is Ref.~\cite{moretti}. Here, scans over the five MSUGRA parameters  were carried out
(for all processes of the type squark-squark-Higgs, though in this 
paper we only concentrate on the  $\tilde{t}_{1}\tilde{t}^{*}_{1}h^{0}$ case). The parameter space explored was defined as follows:

\noindent $\bullet$ $M_{0} = M_{\frac{1}{2}} = 150$ GeV. (These low stop and gaugino masses were chosen to give low stop and Higgs masses in order to enhance the cross section.)

\noindent $\bullet$ $A_{0}$ was set to $-300,0,300$ GeV, in order to investigate its behaviour as it changes sign.

\noindent $\bullet$  Both values of ${\rm sign}(\mu)$ were investigated.

\noindent $\bullet$  $\tan\beta$ was scanned over values between $0$ and $40$.  

\begin{figure}
\centering
  \includegraphics[width=0.6\textwidth]{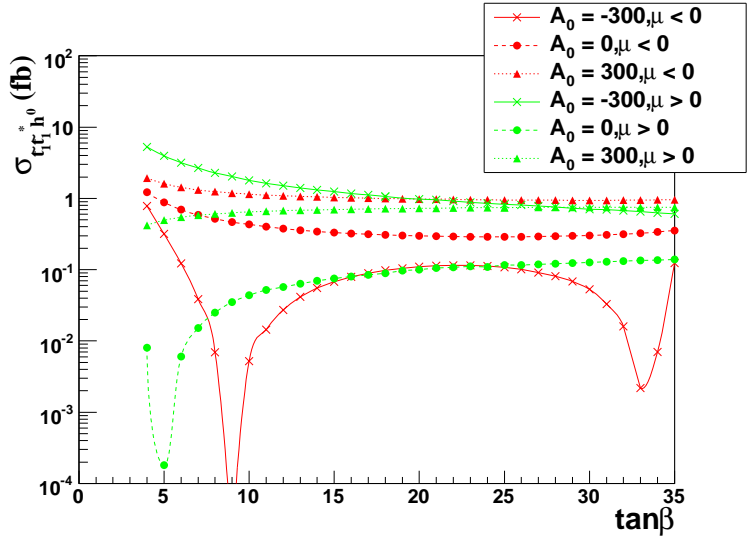}
  \caption{ $\tilde{t}_{1}\tilde{t}^{*}_{1}h^{0}$ cross section values resulting from the variation of MSUGRA parameters.}
 \label{figure:CrossSectionMor}
\end{figure}

Figure \ref{figure:CrossSectionMor} shows the $\tilde{t}_{1}\tilde{t}^{*}_{1}h^{0}$ cross section results from the scans over 
this MSUGRA space. These cross sections are sizably different from those presented in the original paper. 
Different codes were used, however, we have traced back this difference to an error in the convention used 
by \cite{moretti}
in the definition of the stop mixing angle,
$\theta_t$.
While this error is of no consequence for the calculation of the Higgs (the angle enters here through the {\sl square} of
its $\sin$ function) and stop (which is independent by definition) masses, it can have a sizable impact in the
 $\tilde{t}_{1}\tilde{t}^{*}_{1}h^{0}$ coupling, as the latter can be written as 
\begin{equation}
g_{\tilde{t}_{1}\tilde{t}_{1}^* h ^0} = \cos 2\beta \left(\frac{1}{2}\cos^2\theta_t - \frac{2}{3}\sin^{2}\theta_W \cos 2 \theta_t \right) \;\;\;+\;\;\; \frac{m_t^2}{M_Z^2}\;\;\; +\;\;\;\sin 2 \theta_t \frac{m_t}{2M_Z^2}\left(A_t - \mu \cot \beta \right),
\label{equation:CouplingLimit}
\end{equation}
where $M_{W/Z}$ is the mass of the $W/Z$ gauge boson and $\theta_W$ is the weak mixing angle.

The cross sections given in figure \ref{figure:CrossSectionMor} are generally lower than those presented in
 Ref.~\cite{moretti}.  Figure \ref{figure:CrossSectionMor} shows that, in general, the cross section decreases 
with increasing $\tan\beta$. Furthermore, it is  highest for $A_{0} = -300$ GeV and $\mu > 0$.
This is to be expected, as it will give the highest value for $|A_{t}|$, as the negative gauge corrections increase the magnitude of a negative $A_{0}$.
There are also minima in the cross section for two of the scans, which were not seen in the scans presented in \cite{moretti}, and are a result of the new calculation of the stop mixing angle.

In \cite{moretti}, two scans over $M_{\frac{1}{2}}$  were also investigated.
The two scans are defined in table \ref{table:spectrums}, as the ``general low mass spectrum'' and the ``general high mass spectrum''.

\begin{table}[h]
\centering
\begin{tabular}{| c | c | c | c | c | c | }
\hline 
          & $M_0$ (GeV) & $M_{\frac{1}{2}}$ (GeV) & $A_0$ (GeV) & $\tan\beta$ & sign$\mu$\\
\hline
Low mass spectrum  & $200$       & $100$ - $350$         & $-300$      & $35$       & $-1$ \\
\hline
High mass spectrum & $500$      & $100$ - $500$          & $-900$      & $35$       & $-1$ \\
\hline
\end{tabular}
\caption{The MSUGRA parameters corresponding to the high mass and low mass spectra.}
\label{table:spectrums}
\end{table}

\begin{figure} 
  \subfigure[$\tilde{t}_{1}\tilde{t}^{*}_{1}h^{0}$ cross sections for a low mass MSUGRA spectrum.]
{
 \label{figure:M12CrossSectionMor:a}
  \includegraphics[width=0.5\textwidth]{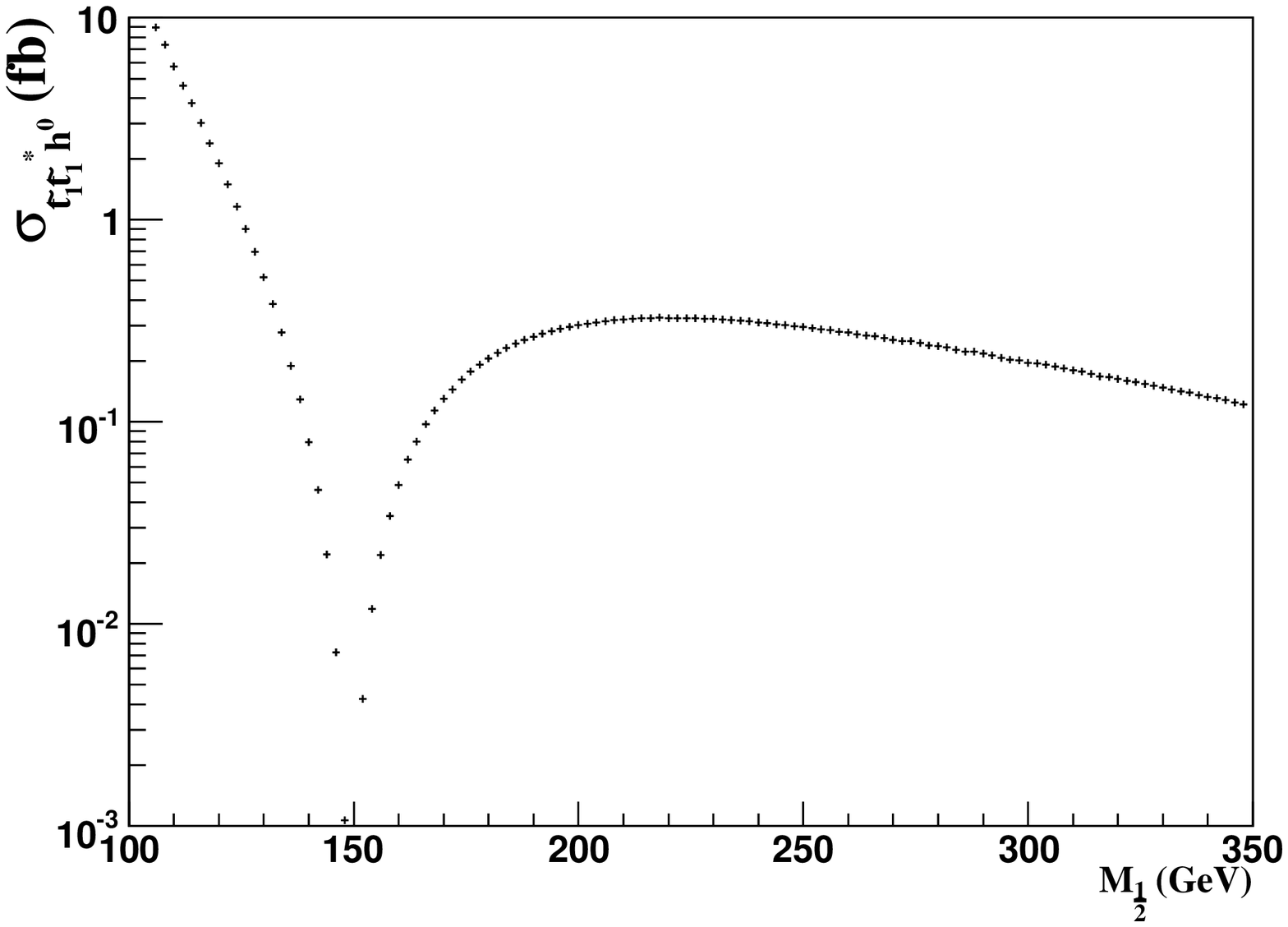}
 }
 \subfigure[$\tilde{t}_{1}\tilde{t}^{*}_{1}h^{0}$ cross sections for a high mass MSUGRA spectrum.]
{
 \label{figure:M12CrossSectionMor:b}
  \includegraphics[width=0.5\textwidth]{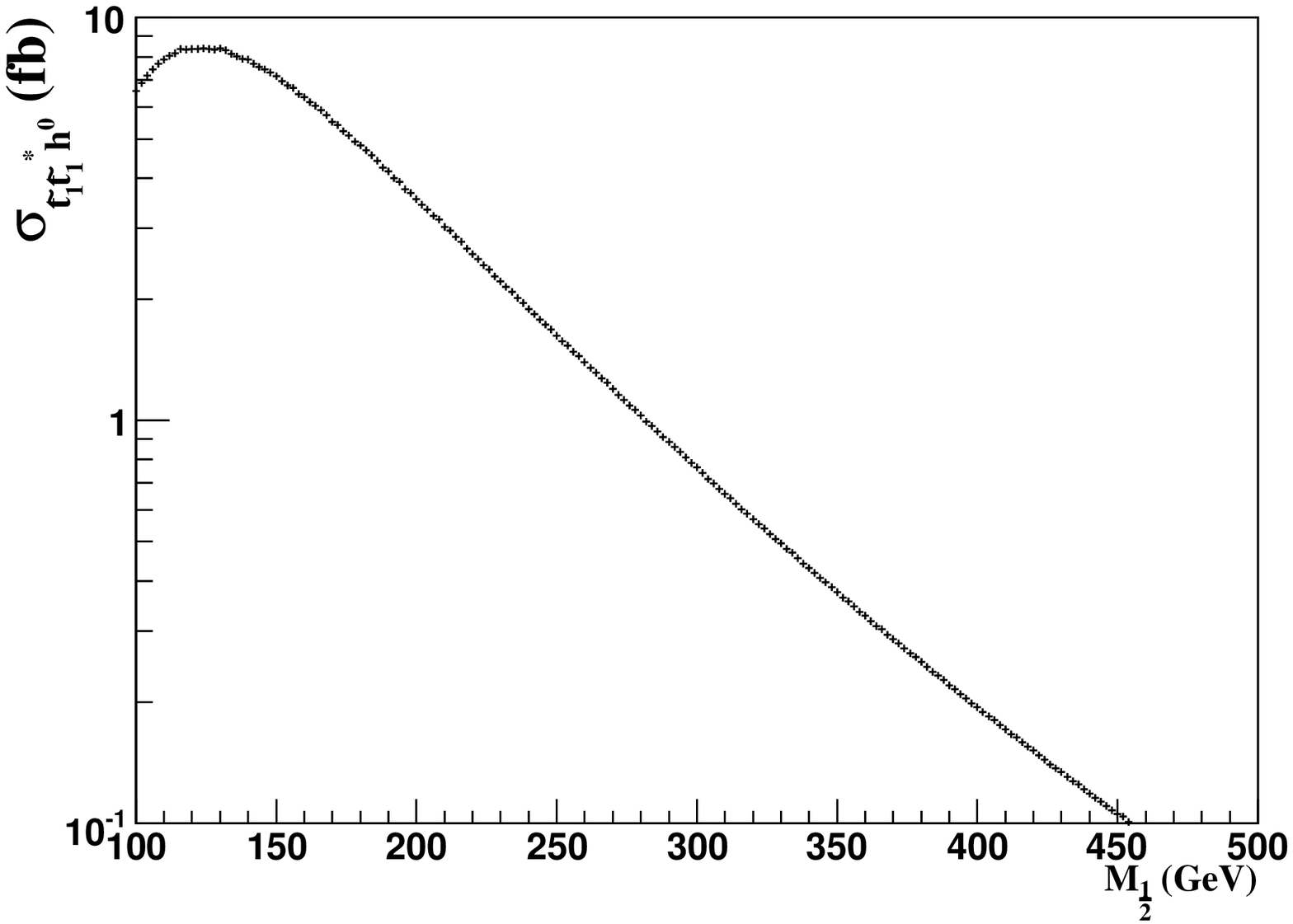}  
}
\caption{$\tilde{t}_{1}\tilde{t}^{*}_{1}h^{0}$ cross section values resulting from the variation of $M_{\frac{1}{2}}$.}
\label{figure:M12CrossSectionMor}

\end{figure}

Figure \ref{figure:M12CrossSectionMor:a} presents the MSUGRA scan for the low mass spectrum whilst figure \ref{figure:M12CrossSectionMor:b} that
for the high mass one.
Figure \ref{figure:M12CrossSectionMor} shows that, in both scans, the cross sections decrease as $M_{\frac{1}{2}}$ increases.
The RGE equations that give the values of the stop and Higgs masses at the EW 
scale include correction terms from gauge interactions.
The latter are dependent on $M_{\frac{1}{2}}$,  giving an 
increase in both the Higgs and stop masses as $M_{\frac{1}{2}}$ 
gets larger.
These higher masses lead to a phase space suppression of $\tilde{t}_{1}\tilde{t}^{*}_{1}h^{0}$
production.
The cross sections calculated here lead to the conclusion that it is unlikely that the $\tilde{t}_{1}\tilde{t}^{*}_{1}h^{0}$ channel will be visible in the MSUGRA regions that were considered in \cite{moretti}.

However, the region of MSUGRA parameter space investigated in Ref.~\cite{djou} is more promising.  
This is characterised, with respect to the region investigated in \cite{moretti}, by a very large negative value of $A_{0}$.
The region investigated is defined in table \ref{table:djouspectrum}.  
The range of $M_0$ values used was chosen to give a range of $m_{\tilde{t}_{1}}$ values that are likely to be detectable at the LHC.

\begin{table}[h]
\centering
\begin{tabular}{| c | c | c | c | c | }
\hline 
$M_0$ (GeV)               & $M_{\frac{1}{2}}$ (GeV) & $A_0$ (GeV) & $\tan\beta$        & ${\rm sign}\mu$\\
\hline
$700$ - $1200$  & $ 300 $               & $-2000$      & $2.5$ and $30$   & $1$ \\
\hline
\end{tabular}
\caption{The region of MSUGRA space investigated in Ref.~\cite{djou}.}
\label{table:djouspectrum}
\end{table}

\begin{figure}
\centering
  \includegraphics[width=0.6\textwidth]{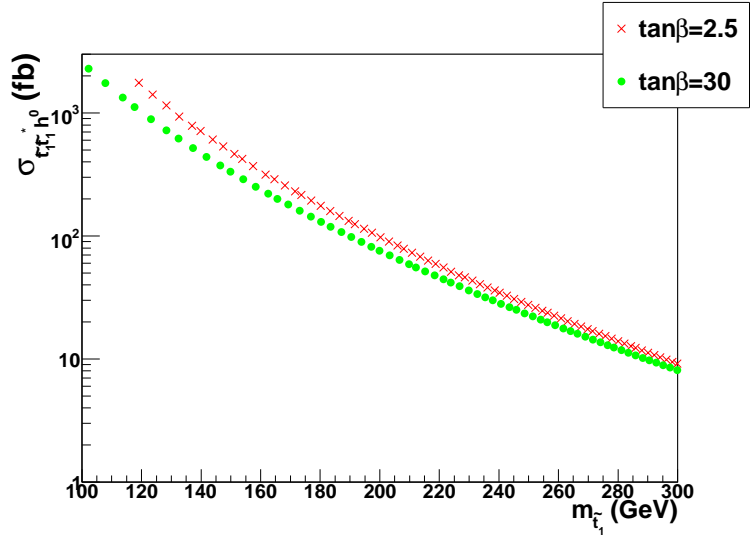}
  \caption{ $\tilde{t}_{1}\tilde{t}^{*}_{1}h^{0}$ cross section values resulting from the variation of $m_{\tilde{t}_{1}}$, over the parameter space defined in table \ref{table:djouspectrum}.}
  \label{figure:CrossSectionDjou}
\end{figure}

Figure \ref{figure:CrossSectionDjou} shows how the $\tilde{t}_{1}\tilde{t}^{*}_{1}h^{0}$ cross section varies with 
$m_{\tilde{t}_{1}}$ over the above area of the MSUGRA parameter space.
The cross section results in figure \ref{figure:CrossSectionDjou} show  rather 
good agreement with those presented in \cite{djou}. As expected, the cross section decreases with decreasing
 $m_{\tilde{t}_{1}}$. Nonetheless, here, the production rates are high enough, so that, for a large range of $m_{\tilde{t}_{1}}$ 
values, they could be visible at the LHC. 
The production rates are not significantly affected by the sign of $\mu$.

\begin{figure}
  \subfigure[$\tilde{t}_{1}$ BRs for $\tan\beta = 2.5$.]
 {
      \label{figure:DjouStopBranching:a}
      \includegraphics[width=0.5\textwidth ]{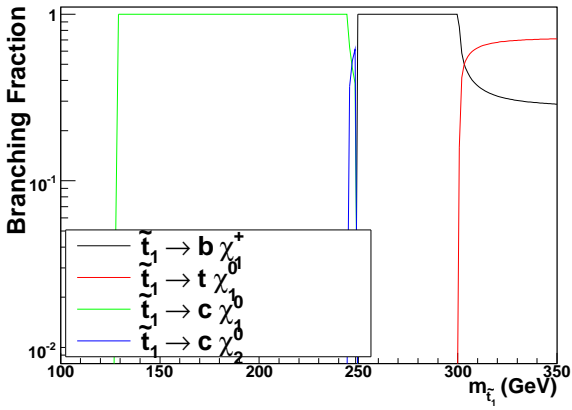}
    }
 \subfigure[$\tilde{t}_{1}$ BRs for $\tan\beta = 30$.]
 {
      \label{figure:DjouStopBranching:b}
      \includegraphics[width=0.5\textwidth ]{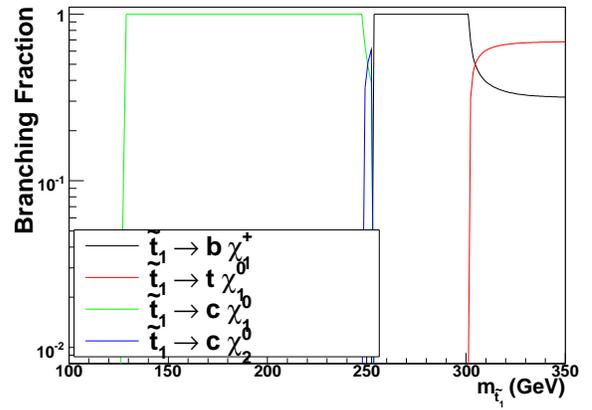}
    }

 \caption{$\tilde{t}_{1}$ BRs resulting from the variation of $m_{\tilde{t}_{1}}$, over the parameter space defined in table \ref{table:djouspectrum}.}
 \label{figure:DjouStopBranching}
\end{figure}

The final state occurring after the decay of the stop quarks and Higgs boson entering
these events also needs to be considered, before drawing any firm conclusions.
The most common decay modes for the stop are to a top quark and a neutralino
(as mentioned already), a bottom quark and a chargino 
or a charm quark and a neutralino.
Figure \ref{figure:DjouStopBranching} shows how the decay modes of the stop change over the range of $m_{\tilde{t}_{1}}$ 
considered in figure \ref{figure:CrossSectionDjou}.
Typical masses for the two lightest neutralinos in this region are $130$ GeV and $250$ GeV, with the lightest chargino also around $250$ GeV.
At $m_{\tilde{t}_{1}} \gtrsim 300$ GeV the stop is most likely to decay to a top quark along with the lightest neutralino.  
For $m_{\tilde{t}_{1}} \lesssim 250$ GeV the most likely decay is to the charm quark and a neutralino. 
A decay to the bottom quark and lightest chargino is most likely for  $ 250$ GeV $\lesssim m_{\tilde{t}_{1}} \lesssim 300$ GeV .
There is also a small area of this parameter space, for  $m_{\tilde{t}_{1}} \sim 250$ GeV, where the stop is most likely to decay to a charm and the second heaviest neutralino, due to a combination of coupling and phase space effects.
For $m_{\tilde{t}_{1}}  \lesssim 130$ GeV, the stop is the Lightest Supersymmetric Particle (LSP) and therefore will not decay
(as long as $R$-parity is conserved). 
The value of $\tan\beta$ has a rather negligible effect on the stop decays.

\begin{figure}
 \centering
 \includegraphics[width=0.6\textwidth]{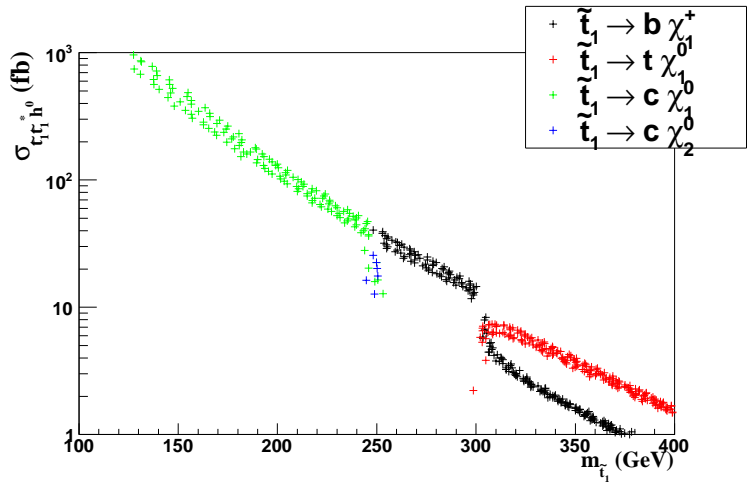}
 \caption{$\sigma_{\tilde{t}_{1}\tilde{t}^{*}_{1}h^{0}}$ for the common stop decay modes, over the parameter space defined in table \ref{table:djouspectrum}.}
 \label{fig:CrossSectionAllBrDjou}
 \end{figure}

 Figure \ref{fig:CrossSectionAllBrDjou} shows the variation in the cross sections for $\tilde{t}_{1}\tilde{t}^{*}_{1}h^{0}$, over the
parameter space defined in table \ref{table:djouspectrum}, for the two stops decaying in each of the common stop decay modes.
 It can be seen from figure \ref{fig:CrossSectionAllBrDjou} that the highest cross section for $\tilde{t}_{1}\tilde{t}^{*}_{1}h^{0}$ occurs when the stop decays to a charm quark and a neutralino. If the  $\tilde{t}_{1}\tilde{t}^{*}_{1}h^{0}$ channel can at all be accessed at the LHC, then 
it is likely that this will occur
in the parameter region where the stop decays to a charm quark and a neutralino.

The most common decay modes for the Higgs boson in the area of parameter space considered in \cite{djou} are those
into bottom or charm quarks or to taus.
The BRs to these decay modes do not vary significantly over the parameter space considered.
This is because the Higgs mass only varies between approximately $97$ and $120$ GeV, much less than the variation in the stop mass.
The decay fraction of the Higgs to bottom quarks is around $85\%$, to taus around $5\%$ and to charm quarks around $3\%$.
The high $h^0 \rightarrow b\bar b$ decay fraction, combined with a $b$-tagging efficiency of around $50\%$, may enable one to extract $h^0$ decays produced at the LHC
in association with stop-antistop pairs.

Having established that the cross sections and BRs are such that it could be possible to access the  $\tilde{t}_{1}\tilde{t}^{*}_{1}h^{0}$ channel at the LHC in the MSUGRA scenario
of Ref.~\cite{djou}, we study an expanded region of parameter space. To this end, additional
scans  were carried out using the adaptive scan program adScan \cite{adscan}.
The adScan program uses adaptive integration algorithms to identify regions of interest in a multidimensional scan.  
It can therefore scan over a wide range, and large number, of parameters faster than a uniform scan. 
The values of the MSUGRA parameters were varied as shown in table \ref{table:msugrascan}.  
These values were chosen to provide a wide Supersymmetric spectrum, but one that is theoretically viable  (i.e., only non-tachyonic particles are present, 
EWSB is achieved, etc.).
Although ${\rm sign}(\mu)$ was initially kept constant and equal to 1, further investigation showed that the value of ${\rm sign}(\mu)$ has little effect on  $\tilde{t}_{1}\tilde{t}^{*}_{1}h^{0}$ production.

\begin{table}[h]
\centering
\begin{tabular}{| c | c | c | c | c | }
\hline 
$M_0$ (GeV)   & $M_{\frac{1}{2}}$ (GeV) & $A_0$ (GeV)       & $\tan\beta$  & sign$(\mu)$\\
\hline
$0$ - $1000$  &   $0$ - $1000$        & $-2500$ - $2500$  & $2$ - $30$  & $1$ \\
\hline
\end{tabular}
\caption{The region of MSUGRA space investigated using the adScan program.}
\label{table:msugrascan}
\end{table}

\begin{figure}
 \centering
\includegraphics[width=0.6\textwidth]{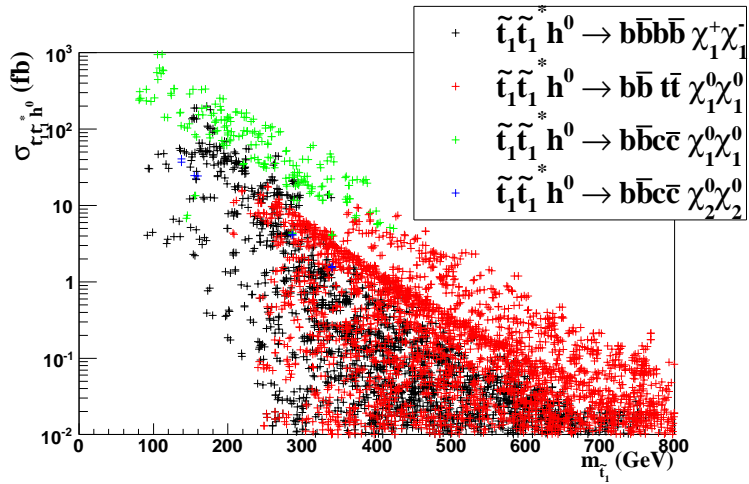}
\caption{ $\tilde{t}_{1}\tilde{t}^{*}_{1}h^{0}$ cross section values resulting from the variation of MSUGRA parameters.}
 \label{figure:MuPlusCrossSection}
\end{figure}

Figure \ref{figure:MuPlusCrossSection} shows how the $\tilde{t}_{1}\tilde{t}^{*}_{1}h^{0}$ cross section varies for this multi-dimensional scan over
the MSUGRA space.
In particular, it shows that the cross sections for $\tilde{t}_{1}\tilde{t}^{*}_{1}h^{0}$ with $\tilde{t}_{1} \rightarrow c \chi^{0}_{1}$ can be 
as high as 1 pb.
For the $\tilde{t}_{1} \rightarrow b \chi^{+}_{1}$ decay channel the cross sections reach a maximum of around 100 fb while  for $\tilde{t}_{1} \rightarrow t \chi^{0}_{1}$ the maximal value is only around 10 fb. 

Production cross sections for the
MSUGRA $\tilde{t}_{1}\tilde{t}^{*}_{1}h^{0}$ signal at the LHC are therefore highest for $\tilde{t}_{1} \rightarrow c \chi^{0}_{1}$ (and charge conjugate) and $h^0\to b\bar b$.

\section*{The $\tilde{t}_{1}\tilde{t}^{*}_{1}h^{0}$ rates in the MSSM parameter space}
While MSUGRA is an appealing representation for a SUSY-breaking scenario, it need not be the one realised in Nature. We have therefore
scanned
over a generic (non-universal) MSSM parameter space, defined at the EW scale, where the important parameters for stop-stop-Higgs hadro-production
are the following.

1. $\tan\beta$, the usual ratio of the VEVs of the Higgs fields.

2. $\mu$, the higgsino mass parameter. 

3. $A_{t}$, the stop trilinear coupling.

4. $m_{\tilde{t}_{L}}$, the left-handed stop mass.

5. $m_{\tilde{t}_{R}}$, the right-handed stop mass.

6. $M_{{A^0}}$, the CP-odd Higgs mass.

\noindent
As the values of these parameters are limited in the MSUGRA scenario, it may be possible that in the general MSSM
 there are areas of parameter space where the $\tilde{t}_{1}\tilde{t}^{*}_{1}h^{0}$ cross section is higher than in MSUGRA.

Multi-dimensional scans over the MSSM parameter space were again carried out using adScan. The values of the MSSM parameters were varied as shown in table \ref{table:mssmscan}.  
Again, as for the MSUGRA scans, values were chosen to provide a wide, but theoretically viable. Supersymmetric spectrum.
\begin{table}[h]
\centering
\begin{tabular}{| c | c | c | c | c | c |}
\hline 
$m_{\tilde{t}_{L}}$ (GeV)  & $m_{\tilde{t}_{R}}$ (GeV) & $A_t$ (GeV)       & $\tan\beta$  & $\mu$ (GeV)            & $m_{{A}}$ (GeV) \\
\hline
$0$ - $1000$             &   $0$ - $1000$          & $-2500$ - $2500$  & $2$ - $30$  &  $-2500$ - $2500$ & $0$ - $1000$  \\
\hline
\end{tabular}
\caption{The region of MSSM space investigated using the adScan program.}
\label{table:mssmscan}
\end{table} 

\begin{figure}
 \centering
\includegraphics[width=0.6\textwidth]{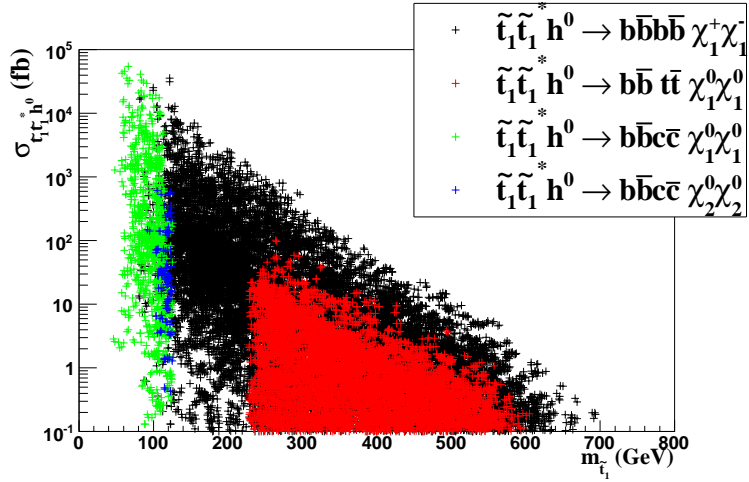}
 \caption{ $\tilde{t}_{1}\tilde{t}^{*}_{1}h^{0}$ cross section values resulting from the variation of MSSM parameters.}
 \label{figure:SUSYCrossSection}
\end{figure}

Figure \ref{figure:SUSYCrossSection} shows how the $\tilde{t}_{1}\tilde{t}^{*}_{1}h^{0}$ cross section varies for this multi-dimensional scan over the
MSSM space.
In particular, it shows that the cross sections for $\tilde{t}_{1}\tilde{t}^{*}_{1}h^{0}$ with $\tilde{t}_{1} \rightarrow c \chi^{0}_{1}$ can be as high as 
100 pb.
For the $\tilde{t}_{1} \rightarrow b \chi^{+}_{1}$ decay channel, the cross sections reach a maximum of around 10 pb and for $\tilde{t}_{1} \rightarrow t \chi^{0}_{1}$ the maximum cross section is around 100 fb. We therefore conclude that within this region of the MSSM parameter space, as for the MSUGRA
study, charm quark decays of stop
quarks provide the highest production cross sections for the $\tilde{t}_{1}\tilde{t}^{*}_{1}h^{0}$ channel at the LHC.

The MSUGRA and MSSM parameter scans above have not enforced experimental limits on the (s)particle masses.
Table \ref{table:ExLimits} shows these for the sparticles contained in the possible $\tilde{t}_{1}\tilde{t}^{*}_{1}h^{0}$ decay chains. 
\begin{table}[h]
\centering
\begin{tabular}{ | c | c | c | c | c | }
\hline
Sparticle & $h^{0}$ & $\tilde{t}_{1}$ & $\tilde{\chi}^{+}_{1}$ & $\tilde{\chi}^{0}_{1}$ \\ 
\hline
Experimental Limit (GeV)  & 93 & 95 & 103 & 39\\
\hline
\end{tabular}
\caption{Experimental limits on the masses of the sparticles relevant to this study.}
\label{table:ExLimits}
\end{table}   
All these limits are taken from LEP2 and Tevatron searches.
A summary of the experimental limits on (s)particle
masses can be found in \cite{pdg}.

In order for the experimental limits to be imposed in a comprehensive way, all sparticle mass limits should be considered, as
each parameter point in the MSSM or MSUGRA space can yield SUSY states not relevant to the $\tilde{t}_{1}\tilde{t}^{*}_{1}h^{0}$ cross section
yet being inconsistent with experiment. Clearly, it is possible to alter the masses of (most) other particles without 
changing the  $\tilde{t}_{1}\tilde{t}^{*}_{1}h^{0}$ cross section.
This can be done by altering the values of any of the MSSM parameters, bar the six that affect the $\tilde{t}_{1}\tilde{t}^{*}_{1}h^{0}$ channel.

\begin{figure}
 \centering
\includegraphics[width=0.6\textwidth]{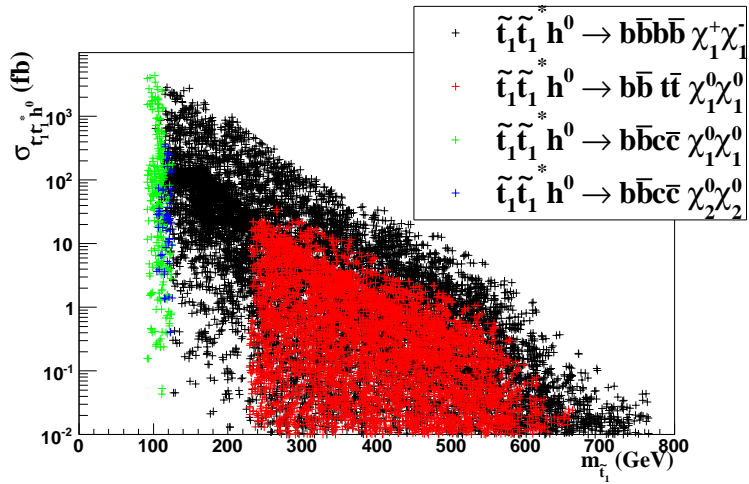}
 \caption{ $\tilde{t}_{1}\tilde{t}^{*}_{1}h^{0}$ cross section values resulting from the variation of MSSM parameters after the experimental constraints in table \ref{table:ExLimits}  have been taken into account.}
 \label{figure:SUSYExCrossSection}
\end{figure}

Performing a scan respecting the experimental limits given in table \ref{table:ExLimits} results in the $\tilde{t}_{1}\tilde{t}^{*}_{1}h^{0}$ cross sections shown in figure \ref{figure:SUSYExCrossSection}.
This  shows that the cross sections for $\tilde{t}_{1}\tilde{t}^{*}_{1}h^{0}$ with $\tilde{t}_{1} \rightarrow c \chi^{0}_{1}$ or $\tilde{t}_{1} \rightarrow b \chi^{+}_{1}$ can reach a few pb.
For the $\tilde{t}_{1} \rightarrow t \chi^{0}_{1}$ decay channel the cross sections reach a maximum of around 50 fb.

\section*{Outlook and summary}
In summary, multi-dimensional scans were carried out over the MSUGRA and MSSM parameter spaces. 
It was found that the highest $\tilde{t}_{1}\tilde{t}^{*}_{1}h^{0}$ cross sections correspond to low stop masses.
In these areas the stop is most likely to decay to a charm quark and the lightest neutralino. Considering that the $h^0$ almost always yields $b\bar b$ pairs, 
the complete final state in presence of these decays will be two charm quarks, two bottom quarks and missing energy.
The final state for the areas where the stop decays to a bottom quark will be four bottom quarks, missing energy, plus two pairs of jets and/or leptons from the chargino decays.
The same final state will occur for the stop to top quark decay, with the additional jets or leptons originating from the $W$ boson produced in the top decay.
There is also the possibility of mixed stop decays, leading to even more variation in the final states. 
The diversity in the final states means that different triggers, backgrounds and selection algorithms will need to be considered for each decay channel. 
Further detailed studies will therefore be required in order to ascertain the visibility of the $\tilde{t}_{1}\tilde{t}^{*}_{1}h^{0}$ channel at the LHC.


\end{document}